\documentclass[twocolumn]{aastex62}
\hypersetup{breaklinks}
\usepackage{needspace}
\usepackage{amsmath}

\submitjournal{ApJ}

\newcommand{\thetitle}{An upper limit on the strength of the extragalactic magnetic field from ultra-high-energy cosmic-ray anisotropy}

\shorttitle{Upper limit on the extragalactic magnetic field} % up to 44 chars
\shortauthors{Bray \& Scaife}

\newcommand{\secref}[1]{Section~\ref{#1}}
\newcommand{\secrefii}[2]{Sections~\ref{#1} and~\ref{#2}}
\newcommand{\eqnref}[1]{Equation~\ref{#1}}
\newcommand{\eqnrefii}[2]{Equations~\ref{#1} and~\ref{#2}}
\newcommand{\figref}[1]{Figure~\ref{#1}}

\newcommand{\tabref}[1]{Table~\ref{#1}}

\let\oldsection\section
\renewcommand{\section}[1]{\Needspace{5\baselineskip} \oldsection{#1} }
\let\oldsubsection\subsection
\renewcommand{\subsection}[1]{\Needspace{5\baselineskip} \oldsubsection{#1} }
\let\oldsubsubsection\subsubsection
\renewcommand{\subsubsection}[1]{\Needspace{5\baselineskip} \oldsubsubsection{#1} }

\newcommand{\fermi}{\mbox{\textit{Fermi}-LAT}}
\newcommand{\swift}{\mbox{\textit{Swift}-BAT}}
\newcommand{\manchester}{JBCA, School of Physics \& Astronomy, University of Manchester, Manchester M13 9PL, UK}
\newcommand{\bfrac}[2]{\ensuremath{\left( \dfrac{#1}{#2} \right)}}

\begin{document}

\title{\MakeUppercase{\thetitle}}

\author{J.\ D.\ Bray}
\affil{\manchester}

\author{A.\ M.\ M.\ Scaife}
\affil{\manchester}

\correspondingauthor{J.\ D.\ Bray}
\email{justin.bray@manchester.ac.uk}

\begin{abstract} % up to 250 words
 If ultra-high-energy cosmic rays originate from extragalactic sources, the offsets of their arrival directions from these sources imply an upper limit on the strength of the extragalactic magnetic field.  The Pierre Auger Collaboration has recently reported that anisotropy in the arrival directions of cosmic rays is correlated with several types of extragalactic objects.  If these cosmic rays originate from these objects, they imply a limit on the extragalactic magnetic field strength of $B < 0.7$--$2.2 \times 10^{-9} \left( \lambda_B / {\rm 1~Mpc} \right)^{-1/2}$~G for coherence lengths $\lambda_B < 100$~Mpc and $B < 0.7$--$2.2 \times 10^{-10}$~G at larger scales.  This is comparable to existing upper limits at $\lambda_B = 1$~Mpc, and improves on them by a factor~\mbox{4--12} at larger scales.  The principal source of uncertainty in our results is the unknown cosmic-ray composition.
\end{abstract}

\keywords{astroparticle physics --- cosmic rays --- magnetic fields}

\section{Introduction}

Magnetic fields are a pervasive ingredient of astrophysical structure, from the small-scale inhomogeneities associated with star formation to the large-scale over-densities associated with galaxy clusters, filaments and the cosmic web.  Across this wide range of scales, a common paradigm is accepted of amplification via dynamo and compression processes; however, in each case the amplification requires the presence of a pre-existing seed field.  The origin of this seed field remains an open question in astrophysics.

One of the key issues with tracing modern-day magnetic fields back to their origin is the problem of saturation effects, which result in amplified field strengths largely independent of their initial values.  Since amplification is linked to local density, it is therefore the least-dense environments that retain the most information about their seed magnetic fields, and are of the greatest value in determining their origin.  The low-density intergalactic medium, which incorporates the voids in the web of large-scale structure, is of particular interest for the study of cosmic magnetism.

There are a variety of mechanisms for placing observational constraints on the strength of the extragalactic magnetic field (EGMF) in voids.  A limit of $B < 9 \times 10^{-10}$~G has been found using power-spectrum analyses of the cosmic microwave background \citep[CMB;][]{ade2014,ade2016}.  An observed absence of correlation between diffuse synchrotron emission and large-scale structure has been used to place a limit on the field strength in filaments which implies a similar limit in voids, $B < 10^{-9}$~G \citep{brown2017}.  These upper limits complement the lower limit of $B \geq 3 \times 10^{-16}$~G set by the non-detection of gamma-ray cascades \citep{neronov2010}.  For a comprehensive review of observational constraints on the EGMF we refer the reader to \citet{durrer2013}.

Another method for probing the EGMF is through observations of ultra-high-energy cosmic rays (UHECRs).  The trajectories of these charged particles are deflected as they pass through magnetic fields, and the magnitude of this deflection acts in principle as a measure of the field strength.  If a UHECR source can be identified and shown to be extragalactic, then the displacement between the source and the corresponding UHECRs depends on the strength of the EGMF.  To date, no individual UHECR source has been conclusively identified, but recent results from the Pierre Auger Observatory show collective correlations of UHECR arrival directions with several types of extragalactic objects, indicating an extragalactic origin for these particles \citep{aab2017,aab2018a}.

In the absence of intervening magnetic fields, we would expect the observed arrival directions of UHECRs to be aligned with their sources within the instrumental resolution.  In practice, there will be an offset due to magnetic deflection of UHECRs by a combination of the EGMF and the Galactic magnetic field (GMF).  The uncertain Galactic component may be neglected, and the entire deflection attributed to the EGMF, in order to place a conservative upper limit on the EGMF contribution.

The concept of constraining the EGMF with this approach has been discussed by \citet{lee1995}, prior to the recent detection of UHECR anisotropy.  More recently there have been detailed analyses of UHECR diffusion in theoretically-motivated models of the EGMF \citep{vazza2017,hackstein2017}, which are able to reproduce the observed large-scale anisotropy, though not yet to discriminate between these models.  There is a clear need for the refinement and application of the approach of \citeauthor{lee1995}, with data from recent UHECR observations, to constrain the EGMF in a simple parameterized model.

In the following, we consider the scenario in which the UHECR anisotropy observed by the Pierre Auger Observatory is associated with one or more of the types of extragalactic objects with which they report correlations, and derive a conditional limit on the strength of the EGMF in the nearby Universe.  In \secref{sec:uhecr} we describe the propagation of UHECRs in the presence of a magnetic field, in \secref{sec:derivation} we derive a new limit on the strength of the EGMF, in \secref{sec:discussion} we discuss this limit in the context of existing constraints, and in \secref{sec:conclusions} we draw our conclusions.

\section{Propagation of ultra-high-energy cosmic rays}
\label{sec:uhecr}

UHECRs are charged particles, consisting of fully-ionized atomic nuclei, and consequently are deflected by magnetic fields.  The effect of these deflections on the propagation of UHECRs depends on the field strength.  For strong magnetic fields, the propagation is fully diffusive, and the arrival direction of a UHECR bears no relation to the direction of its source.  This scenario predicts that the UHECR sky should be primarily isotropic, though with a small degree of anisotropy from the Compton-Getting effect \citep{compton1935}.  For weak magnetic fields, the arrival directions of UHECRs will be offset from their sources by an angle depending on the magnitude of the deflection, which in the small-angle limit is proportional to the field strength.  Any observed correlations of UHECRs with the directions of sources, if such are identified, imply that we are in the latter regime, with the offset angles providing a measure of the magnetic field strength.

In the weak-field scenario, the offset angles between the arrival directions of UHECRs and their sources due to deflections in the EGMF will depend both on the strength of the EGMF and on the scale of its coherent structure.  In general, we expect the EGMF to have a turbulence spectrum that spans a range of scales.  We will consider here two special cases: one in which the coherence length $\lambda_B$ of the EGMF is longer than the distance $D$ to a UHECR source (\secref{sec:uniform}), making it uniform on this scale; and one in which the EGMF consists of independent cells of size ${\lambda_B \ll D}$ (\secref{sec:turbulent}).

\subsection{In a uniform magnetic field}
\label{sec:uniform}

\begin{figure}
 \includegraphics[width=\linewidth]{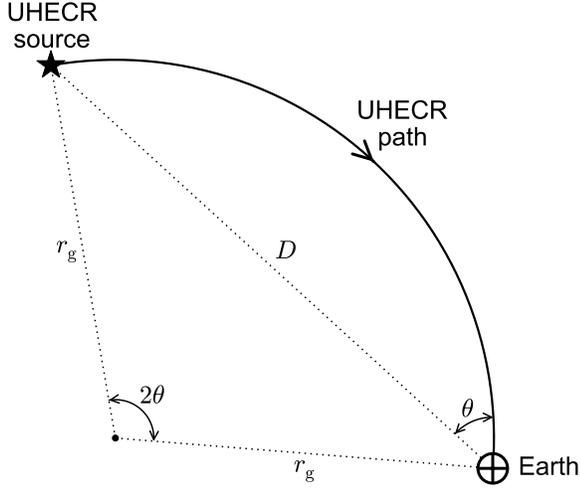}
 \caption{Motion of a UHECR in a uniform magnetic field.  The magnetic deflection of the UHECR causes its arrival direction at Earth to be offset by an angle $\theta$ from the position of its source.  The distance $D$ to the source and the gyroradius $r_{\rm g}$ of the UHECR obey the relation ${D = 2 r_{\rm g} \sin\theta}$.}
 \label{fig:coherent}
\end{figure}

If the EGMF has a coherence length $\lambda_B$ longer than the distance $D$ to a source of UHECRs, a UHECR propagating from this source to Earth will experience a near-uniform magnetic field.  Assuming this field to have a strength $B_\perp$ perpendicular to the motion of the UHECR, it will follow a curved path with a gyroradius
 \begin{equation}
  r_{\rm g} = \frac{E}{Z e c B_\perp} \label{eqn:r_g}
 \end{equation}
where $E$ is the energy of the UHECR, $Z$ its atomic number, $e$ the electron charge, and $c$ the speed of light.  As illustrated in \figref{fig:coherent}, this leads to an offset $\theta$ between the observed arrival direction of the UHECR and the position of its source.  From \eqnref{eqn:r_g} and geometrical considerations, this offset angle can be found as
 \begin{align}
  \sin\theta
   &= \frac{D}{2}\frac{ Z e c B_\perp }{E}
    \label{eqn:theta} \\
   &= 2.65^\circ Z \bfrac{D}{\rm 10~Mpc} \bfrac{B_\perp}{10^{-9}~{\rm G}} \bfrac{E}{10^{20}~{\rm eV}}^{\!\!-1}
    \label{eqn:thetavals}
   .
 \end{align}
Note that this offset angle differs from the deflection of the path of the UHECR as given in equation~(5) of \citet{lee1995} and equation~(135) of \citet{durrer2013}, which is $2\theta$ in our notation.

Given a constraint ${\theta < \theta_{\rm max}}$ on the offset angle due to magnetic deflection by the EGMF, it is possible to place an upper limit on the EGMF strength $B$.  As $B_\perp$ represents the strength of the magnetic field in only two spatial dimensions, and assuming no preferred orientation of the field relative to Earth, we can estimate ${B = B_\perp \sqrt{3/2}}$.  Consequently we obtain the limit
 \begin{align}
  B
   &< \sin(\theta_{\rm max}) \frac{\sqrt{6}}{D} \frac{E}{Z e c}
    \label{eqn:B1} \\
   &< 2.65 \times 10^{-8}~{\rm G} \, \frac{\sin(\theta_{\rm max})}{Z} \bfrac{D}{\rm 10~Mpc}^{\!\!-1} \bfrac{E}{10^{20}~{\rm eV}}
   \label{eqn:B1vals}
  .
 \end{align}

For a uniform magnetic field, UHECRs from different points on the sky will experience a similar deflection, expressed as a rotation around an axis aligned with the local orientation of the EGMF.  The angle $\theta$ can therefore be interpreted as the offset of UHECR arrival directions for a single source, as described above, or the collective offset for a population of sources at a common distance $D$.%  If this population is distributed over the entire sky then the requirement on the coherence length strictly becomes ${\lambda_B > 2 D}$, as the magnetic field must be uniform over the volume containing the entire population.

\subsection{In a turbulent magnetic field}
\label{sec:turbulent}

\begin{figure}
 \includegraphics[width=\linewidth]{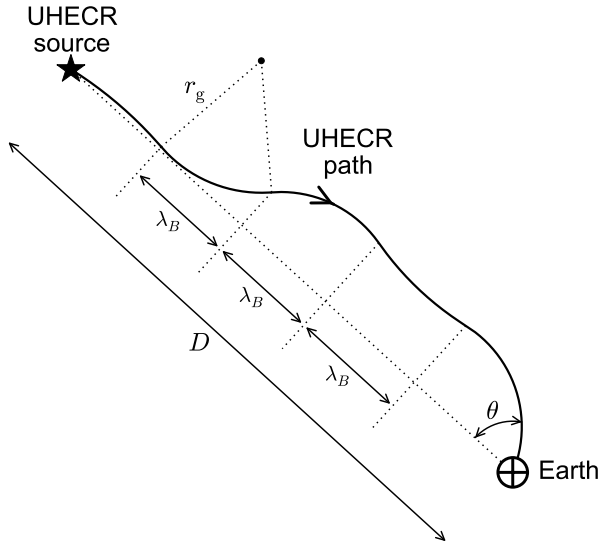}
 \caption{Motion of a UHECR in a turbulent magnetic field with coherence length $\lambda_B$.  A series of small deflections in individual turbulence cells, each approximated as having a uniform magnetic field, leads to an accumulated offset in the UHECR arrival direction ${\theta \propto \sqrt{D/\lambda_B}}$.}
 \label{fig:incoherent}
\end{figure}

If the EGMF is turbulent on small scales --- that is, its coherence length $\lambda_B$ is smaller than the distance $D$ to a source of UHECRs --- then a UHECR from this source will not follow a simple path as shown in \figref{fig:coherent}.  In the limit ${\lambda_B \ll D}$, it will stochastically accumulate a series of small deflections as shown in \figref{fig:incoherent}.  UHECRs from a single source will undergo different deflections, and the source will appear to be smeared out, with a root-mean-square scale
 \begin{align}
  \theta_{\rm rms}
   &\approx \frac{\sqrt{D \, \lambda_B}}{2} \, \frac{Z e c B_\perp}{E}
    \label{eqn:thetarms} \\
   &
   \begin{aligned}
    \approx 0.84^\circ Z \bfrac{D}{\rm 10~Mpc}^{\!\frac{1}{2}} \bfrac{\lambda_B}{\rm 1~Mpc}^{\!\frac{1}{2}}
     \\ \times \bfrac{B_\perp}{10^{-9}~{\rm G}} \bfrac{E}{10^{20}~{\rm eV}}^{\!\!-1}
    . \label{eqn:thetarmsvals}
   \end{aligned}
 \end{align}
If we can place a constraint ${\theta_{\rm rms} < \theta_{\rm max}}$ on this angle then, similarly to \eqnref{eqn:B1}, we can constrain the strength of the EGMF to be
 \begin{align}
  B
   &\lesssim \theta_{\rm max} \frac{\sqrt{6}}{\sqrt{D \, \lambda_B}} \, \frac{E}{Z e c}
    \label{eqn:B2} \\
   &
   \begin{aligned}
    \lesssim 8.37 \times 10^{-8}~{\rm G} \, \frac{\theta_{\rm max}}{Z} \bfrac{D}{\rm 10~Mpc}^{\!\!-\frac{1}{2}}
    \\ \times \bfrac{\lambda_B}{\rm 1~Mpc}^{\!\!-\frac{1}{2}} \bfrac{E}{10^{20}~{\rm eV}}
    . \label{eqn:B2vals}
   \end{aligned}
 \end{align}
Such a constraint may be obtained by observing a smeared-out UHECR source, or the angular scale of a statistical correlation between such sources and UHECR arrival directions.  More generally, observing any structure in the all-sky distribution of UHECRs would imply ${\theta_{\rm rms} \lesssim 1}$~rad.  This limit might be slightly exceeded, at the cost of reducing the amplitude of the observed structure, but in this case the small-angle approximation inherent to \eqnref{eqn:thetarms} breaks down and a more general simulation is required \citep[e.g.][]{vazza2017,hackstein2017}.

\subsection{Other propagation effects}
\label{sec:propagation}

In the preceding discussion we have assumed that the energy and charge of a deflected UHECR remain unchanged as they propagate through the EGMF.  In practice, UHECRs suffer energy losses or attenuation through interactions with background photon fields, so their observed energy on arrival at Earth does not accurately reflect their gyroradius during propagation.  The principal energy-loss mechanisms affecting UHECR protons are pair production \citep[the Bethe-Heitler process;][]{bethe1934} and the photopion interactions responsible for the GZK limit \citep{greisen1966,zatsepin1966}.

The relative impacts of these processes vary depending on energy.  The GZK limit imposes a strong cut-off for UHECR protons with energies exceeding ${E_{\rm GZK} \sim 5 \times 10^{19}}$~eV, which will have a mean free path that decreases to $\lambda_{\rm GZK} \sim 6$~Mpc at $\gtrsim 10^{20}$~eV, but at energies $\lesssim 10^{19}$~eV the GZK limit is effectively equivalent to the cosmological horizon \citep{ruffini2016}.

In contrast to GZK photopion interactions, the Bethe-Heitler pair-production process only removes a small fraction of the energy of a UHECR.  Therefore, although the process has a much shorter mean free path of $\lambda_{\rm BH} \sim 437$~kpc \citep{ruffini2016}, UHECRs will scatter many times before losing a significant portion of their energy.  The horizon imposed by the Bethe-Heitler process is instead defined by the mean energy-loss distance,
% \begin{equation}
%  \lambda_{\rm MEL} = \left( \frac{1}{E}\frac{{\rm d}E}{c{\rm d}t} \right)^{-1},
% \end{equation}
corresponding to the distance at which the energy of the UHECR has fallen to $1/e$ of its original value.  For the Bethe-Heitler process this distance is $\gtrsim 1$~Gpc \citep{ruffini2016}, which sets the effective horizon for UHECR protons with energies less than $E_{\rm GZK}$.

Heavier UHECRs such as iron nuclei can additionally interact with background photons through photo-disintegration, splitting them into lighter nuclei.  This process also imposes a GZK limit, at a similar threshold as photopion interactions do for UHECR protons.  At energies over $E_{\rm GZK}$ the mean free path is ${\lesssim 1}$~Mpc, but at lower energies it is much longer, $\gtrsim 100$~Mpc at $10^{19}$~eV and $\gtrsim 1$~Gpc at $10^{18}$~eV \citep{allard2008}.  Furthermore, to first order photo-disintegration does not change the charge-to-mass ratio (or charge-to-energy ratio), which determines the gyroradius, so the assumption that the gyroradius is constant will approximately hold over distances that exceed this length by a small factor. % Arguably, the value we should use here is the mean energy-loss distance in energy per nucleon, but I can't find this clearly reported.

\section{Derivation of a limit on the extragalactic magnetic field}
\label{sec:derivation}

The Pierre Auger Collaboration has recently reported correlations between UHECR arrival directions and several types of extragalactic objects \citep{aab2017,aab2018a}.  Each of these correlations, if it represents a true association between UHECRs and their sources, implies a limit on the strength of the EGMF.  Per \secref{sec:uniform} and \eqnref{eqn:B1vals}, the offset angles between the UHECR arrival directions and the associated sources imply a limit on any component of the EGMF with a scale larger than the distance to these sources.  Per \secref{sec:turbulent} and \eqnref{eqn:B2vals}, these offset angles also imply a scale-dependent limit on any turbulent component of the EGMF with a coherence length shorter than this distance.

In practice, deflections of UHECRs from extragalactic sources will result from both the GMF and the EGMF.  In general, the GMF component of the deflection will add to that applied by the EGMF, and so attributing the entire deflection to the latter, as we do here, will result in a conservative upper limit on its strength.

\subsection{Dipolar anisotropy of the ultra-high-energy cosmic-ray background}
\label{sec:dipole}

The first element of anisotropy recently detected by the Pierre Auger Observatory in the arrival directions of UHECRs corresponds to a dipole with amplitude 6.5\% and significance ${5.2\sigma}$, in a sample of ${3 \times 10^4}$ events with energies above a threshold of 8~EeV \citep{aab2017}.  The dipole is centered on ($l$,$b$) = ($233^\circ$,$-13^\circ$), with an uncertainty around $\pm 10^\circ$.  This position is separated by $125^\circ$ from the Galactic Center, strongly suggesting an extragalactic origin for these particles, in which case they will have experienced deflections in the EGMF.

\citeauthor{aab2017}\ compare this result with anisotropy in the 2~Micron All-Sky Redshift Survey \citep[2MRS;][]{erdogdu2006}.  The 2MRS recorded the redshifts of 23,000 galaxies selected by their near-infrared flux, which is a good tracer of mass.  Any extragalactic source of UHECRs is likely to have a distribution close to that of matter in the nearby Universe, so this represents a general prediction for the distribution of such sources.  The simplest possible comparison is between the dipole anisotropy in UHECR arrival directions and the dipole moment in the all-sky distribution of 2MRS galaxies.

The flux-weighted dipole in the 2MRS, excluding objects in the Local Group, is centered on Galactic coordinates ($l$,$b$) = ($251^\circ$,$+38^\circ$), with a magnitude defined by the peculiar velocity 1577~km\,s$^{-1}$ \citep{erdogdu2006}.  For Local Group objects only, the dipole is in the direction ($l$,$b$) = ($121^\circ$,$-22^\circ$) with peculiar velocity 220~km\,s$^{-1}$.  Combining these, we find the total dipole to be in the direction ($l$,$b$) = ($243^\circ$,$+38^\circ$), with an uncertainty that will be dominated by that of the first component (${\pm 10^\circ}$). % (\pm 12 deg in l, \pm 10 deg in b; but 12 deg in l is ~10 deg on the sky)

The offset angle between this 2MRS dipole and the UHECR dipole reported by \citeauthor{aab2017}\ is $52 \pm 14^\circ$.  This is sufficiently large to permit a chance coincidence, so it does not, on its own, constitute strong evidence that these anisotropies are associated with one another.  It is possible that the UHECRs responsible for the anisotropy originate from a population of extragalactic objects that is not associated with the distribution of matter in the nearby Universe as measured by the 2MRS.  Alternatively, the UHECRs responsible for the anisotropy may indeed originate from extragalactic objects associated with the 2MRS dipole, and the offset angle may result from their deflection in the GMF or in the EGMF.  The direction of the offset matches that expected from deflections in the GMF \citep{aab2017,jansson2012a}, consistent with this picture, although uncertainties in the composition of UHECRs make it difficult to predict its magnitude.

\subsubsection{Resulting limit on the extragalactic magnetic field}
\label{sec:dipole_lim}

If there is a real association between the nearby-galaxy dipole measured by the 2MRS and the UHECR dipole measured by the Pierre Auger Observatory, it implies a limit on the strength of the EGMF.  The strength of any component of the EGMF with a coherence length larger than the typical distance to the 2MRS sources is constrained by \eqnref{eqn:B1vals}, with ${\theta_{\rm max} = 52 \pm 14^\circ}$ the offset angle between the two dipoles.  The strength of smaller-scale turbulence on the EGMF is constrained by \eqnref{eqn:B2vals}, with ${\theta_{\rm max} = 1~{\rm rad} = 57^\circ}$ required to allow the dipole structure to persist.

Other parameters are required by \eqnrefii{eqn:B1vals}{eqn:B2vals}:
 \begin{itemize}
  \item [$Z$,] the mean atomic number of the UHECRs associated with the dipole.  The composition of UHECRs, in terms of the relative fractions of different elements, is poorly understood, leading to a substantial uncertainty in this value.  Current results in this energy range exclude a composition solely of hydrogen, of heavy nuclei such as iron, or of a mixture of the two, suggesting instead a mixed composition of intermediate elements with likely values in the range $1.7 < Z < 5$ \citep{aab2014,aab2017}. % The LOFAR composition results (buitink2016) suggest a composition of protons or helium (Z = 1 or 2), but are at an energy ~30x lower.  They do disagree with Auger results at the same energy, though.
  \item [$D$,] the typical distance to 2MRS sources responsible for the dipole.  Given the median redshift of sources in the 2MRS of ${z \approx 0.02}$ \citep{erdogdu2006} and a Hubble constant of $H_0 = 67.6$~km\,s$^{-1}$\,Mpc$^{-1}$ \citep{grieb2017}, the median distance is ${D = z c H_0 \approx 90}$~Mpc.  However, we instead take the value ${D = 70}$~Mpc, incorporating the moderate attenuation of UHECRs from the more distant sources \citep{aab2018a}.
  \item [$E$,] the typical energy of UHECRs in the sample above a threshold of 8~EeV.  As the dipole position measures the mean deflection of UHECRs, and these deflections are inversely proportional to energy, we calculate the harmonic mean as the typical value.  Due to the steep spectrum of UHECRs, this value is very close to the threshold: from the modeled spectrum \citep{abraham2010} we calculate it to be ${E = 12}$~EeV with a systematic uncertainty of $\pm 14$\% \citep{aab2015b}.
 \end{itemize}
The parameters for this correlation are listed in \tabref{tab:params}.  Note that the typical energy $E$ is well below the GZK threshold, and $D$ is well within the UHECR horizon at this energy, so the assumption that the UHECR charge/mass ratio remains constant, discussed in \secref{sec:propagation}, approximately holds.

\begin{deluxetable}{lccc}%c}
 \tablewidth{0pt}
 \tablecaption{Parameters for observed UHECR-source correlations \label{tab:params}}
 \tablehead{
   \colhead{Source class}
  &\colhead{$\overline{E}$ (EeV)}
  &\colhead{$\theta_{\rm max}$ ($^\circ$)}
  &\colhead{$D$ (Mpc)}
%  &\colhead{significance}
 }
 \startdata
  2MRS dipole & 12 & $52_{-14}^{+14}$\tablenotemark{a} & \phn70 \\[2pt] % & 5.2$\sigma$ \\[2pt]
  \fermi\ SBGs & 50 & $13_{-3\phn}^{+4\phn}$ & \phn10 \\[2pt] % & 4.0$\sigma$ \\[2pt]
  \fermi\ $\gamma$AGN & \phn75\tablenotemark{b} & \phn$7_{-2\phn}^{+4\phn}$ & \phn150\tablenotemark{b} \\[2pt] % & 2.7$\sigma$ \\[2pt]
  \swift & 50 & $12_{-4\phn}^{+6\phn}$ & \phn70 \\[2pt] % & 3.2$\sigma$ \\[2pt]
  2MRS & 49 & $13_{-4\phn}^{+7\phn}$ & \phn70 \\ % & 2.7$\sigma$ \\
 \enddata
 \tablenotetext{a}{For the turbulent component of the EGMF, the relevant value is ${\theta_{\rm max} = 1~{\rm rad} = 57^\circ}$.}
 \tablenotetext{b}{Due to their high energy and the long distance to the correlated $\gamma$AGN sources, these UHECRs are likely to have been highly attenuated and so are not suitable for deriving a limit on the EGMF.}
 \tablerefs{\citet{aab2017,aab2018a}}
\end{deluxetable}

From the parameters in \tabref{tab:params} and \eqnrefii{eqn:B1vals}{eqn:B2vals}, we derive a scale-dependent limit on the EGMF, under the assumption that the correlation with the 2MRS dipole represents a true association between UHECRs and their sources:
 \begin{align*}
  \frac{B}{\rm G} &<
  \begin{cases}
   1.3\times 10^{-9} \, \bfrac{Z}{2.9}^{\!\!-1} \bfrac{\lambda_B}{\rm 1~Mpc}^{\!\!-\frac{1}{2}} & \lambda_B < 100~{\rm Mpc}\\
   1.3 \times 10^{-10} \, \bfrac{Z}{2.9}^{\!\!-1}  & \lambda_B > 100~{\rm Mpc} .
  \end{cases}
 \end{align*}
% \begin{align}
%  \frac{B}{\rm G} &<
%  \begin{cases}
%   1.3\times 10^{-9} \, \left( \dfrac{Z}{2.9} \right)^{-1} \left( \dfrac{\lambda_B}{\rm 1~Mpc} \right)^{-1/2} & \lambda_B < 100~{\rm Mpc} \\
%   1.3 \times 10^{-10} \, \left( \dfrac{Z}{2.9} \right)^{-1}  & \lambda_B > 100~{\rm Mpc}
%  \end{cases}
% \end{align}
The 100~Mpc scale for the transition between these regimes differs from the scale ${D = 70}$~Mpc because of the introduction of a small-angle approximation from \eqnref{eqn:B1vals} to \eqnref{eqn:B2vals}.  As the GMF may be responsible for some of the observed deflection \citep{aab2017}, a limit incorporating the effect of the GMF may be more stringent than this conservative result.

The principal uncertainty in this result is associated with the range $1.7 < Z < 5$ for the mean atomic number, dominating over smaller uncertainties in the offset angle and energy scale; the nominal values above represent the geometric mean of this range (${Z = 2.9}$).  The possible range for the limit, depending on $Z$, is $B < 0.7$--$2.2 \times 10^{-9} \left( \lambda_B / {\rm 1~Mpc} \right)^{-1/2}$~G for coherence lengths $\lambda_B < 100$~Mpc and $B < 0.7$--$2.2 \times 10^{-10}$~G at larger scales.

\subsection{Intermediate-scale anisotropy of ultra-high-energy cosmic rays}
\label{sec:correlations}

The remaining elements of anisotropy recently detected by the Pierre Auger Observatory are correlations between UHECR arrival directions and extragalactic objects from several catalogs \citep{aab2018a}.  These correlations are on intermediate angular scales (7--13$^\circ$), smaller than the all-sky dipole described in \secref{sec:dipole}, but larger than the resolution of the instrument.  Each correlation represents an excess of UHECRs (above some energy threshold) with arrival directions aligned (within some search radius) with objects in a given catalog, against a null hypothesis of an isotropic distribution of UHECRs.  After imposing a statistical penalty for the \textit{a~posteriori} parameter search, \citeauthor{aab2018a}\ report correlations (with corresponding statistical significances) with sources detected in gamma rays by \fermi\ and classified as starburst galaxies (SBGs; $4.0\sigma$) or active galactic nuclei ($\gamma$AGN; $2.7\sigma$), X-ray sources detected by \swift\ ($3.2\sigma$), and infrared sources detected by the 2MRS ($2.7\sigma$).  In each case, the best fit corresponds to a small fraction (7--16\%) of UHECRs originating from objects of the specified type, and the remainder constituting an isotropic background.  The best-fit energy thresholds are in the range \mbox{38--60}~EeV, and the best-fit search radii of 7--13$^\circ$ define the angular scale of the correlations.

Unlike the result described in \secref{sec:dipole}, these correlations directly associate UHECRs with extragalactic sources.  Barring an unlikely chance coincidence, each correlation implies that some fraction of UHECRs originate from the corresponding type of extragalactic object, or from another source class with a correlated extragalactic distribution.  Note that these results are not fully independent, as the extragalactic objects in each result are correlated with one another; \citeauthor{aab2018a}\ also consider joint fits to multiple source classes, which we neglect here.

In this scenario, the offsets between the arrival directions of UHECRs and their corresponding sources result from the combination of deflections in the GMF and EGMF.  Deflection in a component of the EGMF with a large coherence length would result in a systematic offset of UHECRs from multiple sources in a common direction, but \citeauthor{aab2018a}\ do not say whether such an effect is observed.  The use of a fixed search radius for UHECRs around a prospective source, irrespective of its distance, corresponds to an expectation of Galactic deflections only: for deflections in the EGMF, UHECRs from more distance sources would have a larger offset angle, meriting a larger search radius.

\subsubsection{Resulting limit on the extragalactic magnetic field}
\label{sec:correlations_lim}

For each of the correlations reported by \citet{aab2018a} we represent the typical energy with the harmonic mean $\overline{E}$ above the corresponding best-fit energy threshold, as in \secref{sec:dipole_lim}.  For the typical source distance $D$ we use the radii calculated by \citeauthor{aab2018a}\ within which 90\% of the UHECR flux from the corresponding source population is expected to originate, allowing for attenuation.  These values are listed in \tabref{tab:params}, along with the angular scale of each correlation, which we take as an upper limit $\theta_{\rm max}$ on the offset due to the EGMF, conservatively neglecting any deflection in the GMF.  As in \secref{sec:dipole_lim}, we take the mean atomic number to be in the range ${1.7 < Z < 5}$.

The energies of the UHECRs exhibiting these correlations are substantially higher than those responsible for the dipole anisotropy described in \secref{sec:dipole}, and hence more susceptible to attenuation (see \secref{sec:propagation}).  For the correlations with \fermi\ SBGs, and \swift\ and 2MRS sources, the typical energies do not exceed the GZK threshold, and so propagation with minimal attenuation is likely over the ${D \leq 70}$~Mpc distances involved.  However, the correlation with \fermi\ $\gamma$AGN involves more energetic UHECRs, above the GZK threshold, propagating over longer (${D = 150}$~Mpc) distances, and so this population of UHECRs is likely to have undergone substantial attenuation through the processes described in \secref{sec:propagation}, violating the assumptions behind the calculations in \secrefii{sec:uniform}{sec:turbulent}.  This is consistent with the results of \citet[][Figure~1]{aab2018a}, which show attenuation to have a substantial effect only on the correlation with $\gamma$AGN.  We therefore exclude this specific correlation from further analysis.

It is notable from \tabref{tab:params} that the typical offset angle $\theta_{\rm max}$ has an approximately inverse relation with UHECR energy, as expected from \eqnref{eqn:theta}, but does not increase with the typical distance $D$ to the class of correlated sources.  This is the outcome that would result if the deflection were entirely due to the GMF, and thus irrespective of the distance to the source, whereas deflections due to the EGMF will be greater for more distant sources.  Further examination of this trend may allow discrimination between the GMF and EGMF contributions to the deflection of UHECRs, but the uncertainty in $\theta_{\rm max}$ is too large to permit this with the current data.  For the present, it is safe to say that the limit obtained by attributing the entire deflection to the EGMF, as we do here, is likely to be quite conservative.

For the remaining correlations we calculate limits (summarized in \tabref{tab:lims}) on the strength of the EGMF as in \secref{sec:dipole_lim}, under the assumption that each correlation represents a true association between UHECRs and their sources.  For the correlation with \fermi\ SBGs, the resulting limit is
 \begin{align*}
  \frac{B}{\rm G} &<
  \begin{cases}
   3.3\times 10^{-9} \, \bfrac{Z}{2.9}^{\!\!-1} \bfrac{\lambda_B}{\rm 1~Mpc}^{\!\!-\frac{1}{2}} & \lambda_B < 10~{\rm Mpc}\\
   1.0 \times 10^{-9} \, \bfrac{Z}{2.9}^{\!\!-1}  & \lambda_B > 10~{\rm Mpc} ,
  \end{cases}
 \end{align*}
which is substantially less constraining than the limit in \secref{sec:dipole_lim}, due to the shorter typical distances to the correlated sources.  The limit resulting from the correlation with \swift\ sources is
 \begin{align*}
  \frac{B}{\rm G} &<
  \begin{cases}
   1.1\times 10^{-9} \, \bfrac{Z}{2.9}^{\!\!-1} \bfrac{\lambda_B}{\rm 1~Mpc}^{\!\!-\frac{1}{2}} & \lambda_B < 70~{\rm Mpc}\\
   1.3 \times 10^{-10} \, \bfrac{Z}{2.9}^{\!\!-1}  & \lambda_B > 70~{\rm Mpc}
  \end{cases}
 \end{align*}
and the equivalent limit from the correlation with 2MRS sources is almost identical to this, being only 10\% higher (less constraining).  These two are both similarly close to the limit based on the dipole anisotropy described in \secref{sec:dipole_lim}: with no significant loss of precision compared to the uncertainties in the data, we can regard these three correlations to establish a single limit.

\begin{deluxetable}{lcc}
 \tablewidth{0pt}
 \tablecaption{EGMF limits derived from UHECR-source correlations \label{tab:lims}}
 \tablehead{
   \colhead{Source class}
  &\colhead{$B_{\rm uni} (\frac{Z}{2.9})^{-1}$ (G)}
  &\colhead{$B_{\rm turb} (\frac{Z}{2.9})^{-1} (\frac{\lambda_B}{1~{\rm Mpc}})^{-\frac{1}{2}}$ (G)}
 }
 \startdata
  2MRS dipole\tablenotemark{a} & $\phn1.3 \times 10^{-10}$ & $1.3 \times 10^{-9}$ \\[2pt]
  \fermi\ SBGs                 & $   10.2 \times 10^{-10}$ & $3.3 \times 10^{-9}$ \\[2pt]
  \swift\tablenotemark{a}      & $\phn1.3 \times 10^{-10}$ & $1.1 \times 10^{-9}$ \\[2pt]
  2MRS\tablenotemark{a}        & $\phn1.4 \times 10^{-10}$ & $1.2 \times 10^{-9}$ \\[2pt]
 \enddata
 \tablecomments{$B_{\rm uni}$ represents an upper limit on a uniform component of the EGMF, while $B_{\rm turb}$ represents an upper limit on a turbulent component of the EGMF dependent on its coherence length $\lambda_B$.}
 \tablenotetext{a}{Corresponding limits shown in \figref{fig:limplot}.}
\end{deluxetable}

\section{Discussion}
\label{sec:discussion}

\begin{figure}
 \includegraphics[width=\linewidth]{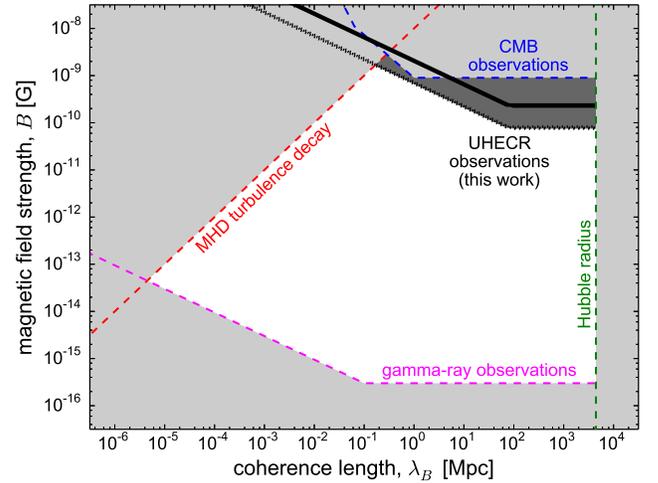}
 \caption{Parameter space for the strength $B$ and coherence length $\lambda_B$ of the EGMF in voids, showing regions excluded by past limits (light shaded) and this work (dark shaded).  The near-identical limits placed in this work, based on UHECR observations, have a substantial uncertainty associated with the mean UHECR atomic number $Z$; solid and dotted lines show respectively the cases $Z=1.7$ and $Z=5$, which represent the range permitted by current composition measurements.  Theoretical constraints are set by MHD turbulence, which causes the decay of short-scale modes in magnetic fields \citep{durrer2013}, and by the Hubble radius, which places an upper limit to the size of any observable structure.  The lower limit is set by the non-detection of gamma-ray cascades \citep{neronov2010}.  The upper limit shown from CMB observations is a projection from \citet{paoletti2011}, as represented by \citet{durrer2013}, and compatible with the limit $B < 9 \times 10^{-10}$~G established with Planck data \citep{ade2016}.}
 \label{fig:limplot}
\end{figure}

At present there are no direct measurements of the strength of the EGMF in voids, but various limits have been established in terms of its strength $B$ and coherence length $\lambda_B$.  The limits derived in this work, based on the correlations of UHECR arrival directions with the dipolar distribution of 2MRS sources and with \swift\ and 2MRS sources on smaller angular scales, are shown in \figref{fig:limplot}.  We also show previous limits, discussed below, confining ourselves for brevity to the most constraining measurements only.  For a more comprehensive review of the observational and theoretical limits on the strength of the EGMF in voids we refer the reader to \citet{durrer2013}.

Using the non-detection of the secondary photon-photon cascade, lower limits on the strength of the EGMF in voids have been set observationally using gamma-ray measurements from \fermi\ \citep{neronov2010,tavecchio2011}.  The general limit given by \citet{neronov2010} is $B \geq 3 \times 10^{-16}$~G, but towards individual blazars the limits span the range $\sim 10^{-17}$--$10^{-14}$~G when considering various emission and suppression scenarios. 

On the smallest scales a theoretical limit is set by the termination of evolutionary tracks in $(B, \lambda_B)$ space for various magnetohydrodynamic (MHD) turbulence scenarios.  On the largest scale a limit is set by the Hubble radius, $\ell_{\rm H}$; fields coherent on scales larger than the Hubble radius are possible if due to seed fields that were generated during inflation.  Observational limits for such fields with coherence lengths longer than the Hubble radius are not measurable and the upper-limit constraint of $B \lesssim 10^{-9}$~G on this scale currently comes from CMB power-spectrum analysis \citep{ade2014,ade2016}.  This upper limit extends uniformly to smaller scales (1~Mpc $\lesssim \lambda_B < \ell_{\rm H}$) and can vary somewhat (within a factor of $\sim 5$) when considering different primordial field scenarios.  The strongest constraint of $B < 9\times 10^{-10}$~G is given by a scenario in which scale-invariant primordial magnetic fields are considered.  These power-spectrum limits are more constraining than those from Faraday rotation of the CMB by several orders of magnitude \citep{ade2016}.  On smaller scales ($\lambda_B \lesssim 1$~Mpc) the behavior of the CMB upper limit becomes more complex as spectral distortions need to be taken into account. 

The results presented here further constrain the upper limits on the magnetic field strength in voids on scales ${\lambda_B > 100}$~Mpc by around an order of magnitude (factor of $\sim 4$--12), depending on the composition of UHECRs.  Composition is the largest source of uncertainty in this limit, as shown in \figref{fig:limplot} and discussed in \secrefii{sec:dipole_lim}{sec:correlations_lim}.

\section{Conclusion}
\label{sec:conclusions}

We have derived an upper limit on the strength of the EGMF, conditional on the distribution of UHECR arrival directions being associated with one or more of the types of extragalactic objects with which correlations have been observed \citep{aab2017,aab2018a}.  Three correlations of UHECR arrival directions --- with an all-sky dipole in the distribution of 2MRS sources, and on smaller angular scales with both \swift\ and 2MRS sources --- each imply a similar limit (within ${\sim 10}$\%).  This implied limit is similar to existing constraints from CMB observations for fields with a coherence length around 1~Mpc, and a factor \mbox{4--12} more constraining for fields with a coherence length ${>100}$~Mpc.

The UHECR dipole has a statistical significance of 5.2$\sigma$, but its correlation with the 2MRS dipole may be a chance alignment, if UHECRs do not originate from a class of object correlated with the extragalactic distribution of mass in the nearby Universe.  The smaller-scale correlations with \swift\ and 2MRS have significances of 3.2$\sigma$ and 2.7$\sigma$ respectively, and are not susceptible to such an alternate explanation.  Our derived limit on the EGMF holds if any of these three correlations represents a true association between UHECRs and their sources; but note that the last two correlations are not completely statistically independent.

These results suggest that techniques for probing cosmic magnetic fields on large scales or amplified fields derived from them, such as observations of diffuse synchrotron or radio polarization, will not achieve a detection until they improve substantially in sensitivity.  Conversely, if such techniques were to achieve a detection, it would cast doubt on the current evidence for an extragalactic origin of UHECRs.  There remains a parameter space spanning five orders of magnitude in EGMF strength between our upper limit and the lower limit established by gamma-ray observations \citep{neronov2010}.

The major source of uncertainty in our limit is the unknown UHECR composition, which is the subject of continued investigation.  The Pierre Auger Observatory is undergoing an upgrade to enable it to discriminate between the muonic and electromagnetic components of particle cascades initiated by UHECRs \citep{aab2016b}, which will improve its ability to discriminate between UHECRs of different elements.  Competitive precision in cosmic-ray composition measurements has also been demonstrated at lower energies by radio measurements with LOFAR \citep{buitink2016}, which may be extended to UHECRs with the upcoming SKA \citep{huege2014}.

In principle, if the strength of the EGMF lies close to the limit established here, it may be possible to detect and measure it using UHECR observations.  This will require improved models of the Galactic magnetic field, so the Galactic contribution to the deflection of UHECRs can be simulated and subtracted \citep{farrar2017}.  It will also benefit from greater signal statistics, which may be accomplished through future instruments with larger collecting areas such as JEM-EUSO \citep{takahashi2009}.  The potential to measure the EGMF through this technique will also depend on a precise knowledge of UHECR composition.

\acknowledgements
The authors thank R.E.~Spencer for helpful comments, and gratefully acknowledge support from ERC-StG\,307215 (LODESTONE).

\bibliographystyle{aasjournal}
\bibliography{all}

\end{document}